# Selecting optimal parallel microchannel configuration(s) for active hot spot mitigation of multicore microprocessors in real time

**Lakshmi Sirisha Maganti** [a, 1], **Purbarun Dhar** [b, 2], **T Sundararajan** [a, $] **and Sarit K Das** [a, b, $]

[a] Department of Mechanical Engineering, Indian Institute of Technology Madras,
Chennai–600036, India

[b] Department of Mechanical Engineering, Indian Institute of Technology Ropar,
Rupnagar–140001, India

E–mail: [1] lakshmisirisha.maganti@gmail.com , [2] purbarun@iitrpr.ac.in

[$] Corresponding author: skdas@iitrpr.ac.in , tsundar@iitm.ac.in

Phone: +91–1881–24–2173

## Abstract

Design of effective micro cooling systems to address the challenges of ever increasing heat flux from microdevices requires deep examination of real time problems and has been tackled in depth. The most common (and apparently misleading) assumption while designing micro cooling systems is that the heat flux generated by the device is uniform, but the reality is far from this. Detailed simulations have been performed by considering non uniform heat load employing the configurations U, I, Z for parallel microchannel systems with water and nanofluids as the coolants. An Intel® Core™ i7–4770 3.40GHz quad core processor has been mimicked using heat load data retrieved from a real microprocessor with non-uniform core activity. The study clearly demonstrates that there is a non-uniform thermal load induced temperature maldistribution along with the already existent flow maldistribution induced temperature maldistribution. The suitable configuration(s) for maximum possible overall heat removal for a hot zone while maximizing the uniformity of cooling have been tabulated. An Eulerian–

Lagrangian model of the nanofluids show that such 'smart' coolants not only reduce the hot spot core temperature, but also the hot spot core region and thermal slip mechanisms of Brownian diffusion and thermophoresis are at the crux of this. The present work conclusively shows that high flow maldistribution leads to high thermal maldistribution, as the common prevalent notion, is no longer valid and existing maldistribution can be effectively utilized to tackle specific hot spot location, making the present study important to the field.

# 1. INTRODUCTION

Thermal management of microelectronic devices has become a challenge to researchers not only due to generation of extremely high heat fluxes but also due to generation of non-uniform heat fluxes from microelectronic devices. Channels with hydraulic diameter of few microns arranged in parallel and connected by inlet and outlet manifolds constitute parallel microchannel cooling systems (PMCS). Such systems have received great attention for cooling high performance microelectronic devices. Tuckerman and Pease [1] proposed a novel concept of employing such high area to volume ratio devices for efficient high heat flux removal from small areas. Sasaki et al. [2] and Kishimoto et al. [3] stressed on design and functioning characteristics of such complex flow domains and those studies can be found in articles. Peng et al. [4] and Judy et al. [5] focused on fundamental understanding of underlying mechanisms in such complex domains and applicability of continuum theory for fluid flow in microchannels. The focus then shifted to practical applicability and challenges that occur in cooling such high heat flux devices using PMCS. Two main challenges which decrease the potential of these complex flow path devices are large pressure drop and non-uniform distribution of fluid (termed as flow maldistribution) among channels. Challenges to be addressed in such complex flow systems in order to achieve high thermal performance in cooling microelectronic devices have been extensively reviewed by Kandlikar [6, 7] and suggestions were proposed.

Kumaraguruparan et al. [8], Siva et al. [9] and Maganti et al. [10] concentrated the studies on understanding the flow distribution among such complex flow domains and the implications of the same on thermal performance of device. Siva et al. [11] reported on the applicability of flow distribution models of macro channels to microchannels and concluded that such models are ineffective for prediction of fluid distribution among microchannels. Then the focus shifted towards effect of flow maldistribution on cooling of microelectronic devices and studies by Hetsroni et al. [12] and Nielsen et al. [13] concluded that fluid maldistribution will induce non-uniform cooling of device, often leading to formation of unintentional hot spots. Since complete eradication of flow maldistribution is not possible, research community emphasized on methods which will improve thermal performance of such devices. Among the several proposed methods, employing efficient heat transfer fluids such as nanofluids is one of the best methods, as reported by Li and Kleinstreuer [14], Escher et al. [15] and Lee and Mudawar [16]. Employing nanofluids as working fluid in such complex flow domains not only improves cooling but also uniformity of cooling (Maganti et al. [17]) because of smart nature of nanofluids at high temperatures, reported by Das et al. [18]. In addition, there is the challenge of designing such effective cooling systems as in reality heat emitted by microelectronic devices will never be uniform and leads to further non-uniform cooling of microelectronic devices (Maganti et al. [17] and Cho et al. [19]).

The present article concentrates on the effects of non-uniform thermal load by mimicking an Intel® Core™ i7–4770 3.40 GHz quad core processor to understand the cooling performance of parallel microchannel cooling systems in real time scenario. The computations involved employ real time heat load data extracted from such a microprocessor working at 70 % of its rated peak load and with preferential usage of its physical cores to amplify the effects of non uniform heat generation. In addition, the change in location of hot spot because of combined effect of flow maldistribution and non-uniform thermal load has been reported and mitigation protocols have been proposed to ensure device safety. It has been clearly shown that the advection of heat from the active heater and the nature of flow system lead to drastic spreading of the heat to regions of the chip which remain relatively cooler when uniform heat flux assumption are considered. Furthermore, to the best of our knowledge, for the first time, suitable configurations have been proposed based on location of active heaters to mitigate hot spot and details have been thoroughly tabulated. The paper also shows the effectiveness of nanofluids in such scenarios as these smart fluids not only reduce the hot spot core temperatures but also shrink the hot spot size (due to localized thermos-fluidic slip mechanisms of the nanoparticles), thereby ensuring enhanced safety of microprocessors.

## 2. Computational formulation and details

The present work investigates the effects of non-uniform thermal load induced temperature maldistribution along with the inevitable flow maldistribution induced temperature maldistribution in cooling electronic devices using parallel micro channel cooling systems (PMCS). The temperature profiles are gathered for real time Intel i7 quad core processor and the optimal microchannel configuration (U, I or Z) is suggested for a particular case of a given active microprocessor core. The calibre of nanofluids to cool the hot spots better as well as to reduce the size of such spots to ensure better thermal safety of device has also been investigated. In order to understand the realistic performance of nanofluids when non-uniform heat load is applied, an Eulerian–Lagrangian Discrete Phase Model (DPM) approach has been employed so as to incorporate the thermo-fluidic slip mechanisms of nanoparticles which have been reported to be of paramount importance in uncovering the real potential of nanofluids computationally (Maganti et al. [10], Maganti et al. [17]).

### 2.1. Governing equations

The governing equations for Eulerian-Lagrangian approach are mass, momentum and energy conservation with the source terms for particle momentum and energy incorporated and the equations solved are expressed as follows

$$\frac{\partial \rho}{\partial t} + \nabla \cdot (\rho \vec{V}) = 0 \tag{1}$$

$$\frac{\partial \rho \vec{V}}{\partial t} + \nabla \cdot (\rho \vec{V} \vec{V}) = -\nabla P + \nabla \cdot \left(\mu(\nabla \vec{V} + \nabla V^T)\right) + S_m \tag{2}$$

$$\partial C \left[\frac{\partial T}{\partial t} + \vec{V} \cdot \nabla T\right] = \nabla \cdot (k \nabla T) + S_e \tag{3}$$

Negligible viscous dissipation and incompressible flow have been assumed. In the above equations ρ is density of continuous phase, V velocity of fluid, t is time, P is pressure, C is specific heat, k is thermal conductivity, $S_m$ is source term representing momentum exchange and $S_e$ represents source term energy exchange between continuous phase and discrete phase. For a Lagrangian system of reference, the governing equation for the motion of the nanoparticles can be expressed based on Newton's second law as

$$\frac{dV_p}{dt} = F_D(V - V_p) + \frac{g(\rho_p - \rho)}{\rho_p} + F \tag{4}$$

Where, $V$ and $V_p$ are velocities of continuous phase and discrete phase respectively and $\rho$ and $\rho_p$ are density of fluid and nanoparticle respectively. $F_D$ is drag force acting on particle and $F$ is the sum of all specific force acting on individual particle. The expression for net force acting on particle given as follows

$$F = F_B + F_T + F_L + F_V + F_p \tag{5}$$

Where, $F_B$ is force due to Brownian fluctuations, $F_T$ is due to thermophoretic drift, $F_L$ is due to Saffman lift, $F_V$ is due to virtual mass and $F_p$ is due to pressure gradient. The expressions for these force components are as reported by the present authors (Maganti et al. [10]).

## 2.2. Simulation details

To understand the non-uniform thermal load induced thermal patterns in microprocessors, a 3-D, parallel microchannel domains of various configurations (U, I and Z) have been created, meshed and the conventional governing equations of mass, momentum and energy are solved using ANSYS Fluent 14.5 solver. The domain dimensions and working fluid details are as follows: channel hydraulic diameter is 100 μm, area ratio ($A_p/A_c$) is 6, number of channels are 15, aspect ratio of channel is 0.1, working fluids are water and alumina-water (5vol. %) nanofluids and Re is 300. The inlet of the microchannel domain is employed to stream the particle phase which then behaves as an independent yet simultaneous entity within the domain as per the governing equations mentioned earlier. Fig. 1(a) shows the geometry employed and Fig. 1(b) shows arrangement of heaters to apply non-uniform thermal load to parallel microchannel cooling system in order to mimic a real quad core processor. From the figure it can be observed that the PMCS have been provided for heat spreader which is located on the top of the processor. In the event of a heat sink with fan unit, there are additional thermal resistances, which in case of computing clusters are often detrimental to the thermal safety of the embedded processor. Accordingly, the present approach lies in removing such additional resistances while simultaneously introducing a better method of active cooling. It is also technically difficult to manufacture cooling system on the processor itself due to design and electrical constraints. The best feasible solution thereby lies in microchannel based cooling of the heat spreader. This aluminium casing is attached directly to the microchannel and hence is the best possible option towards in-situ cooling of hyperactive processors.

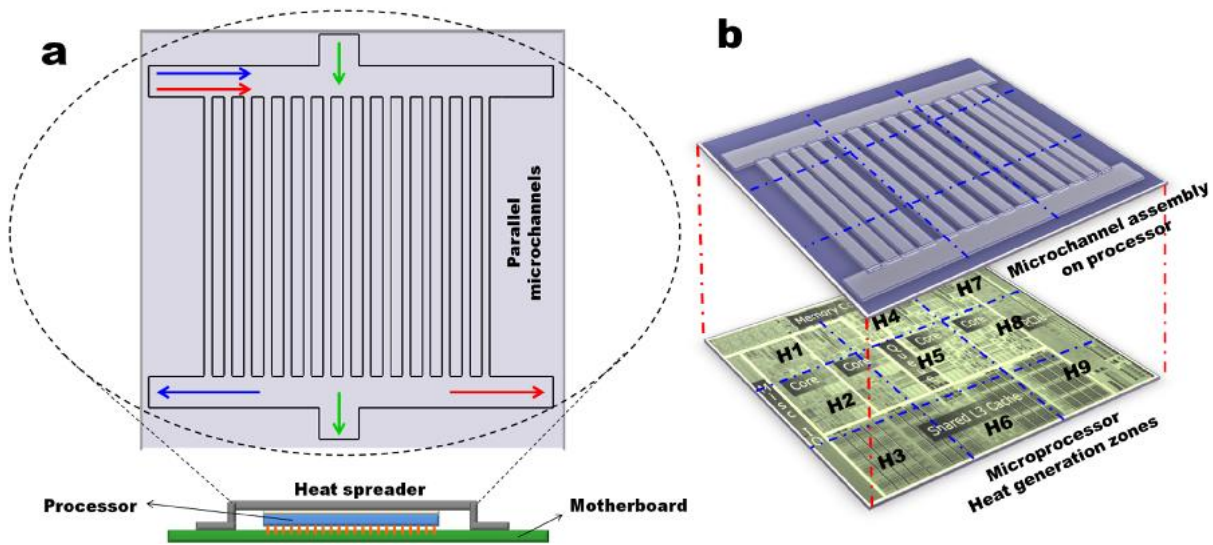

**FIGURE 1:** Exploded view of the computational domain and arrangement of heaters to mimic a real time microprocessor and its non-uniform heat release. The arrangement of components of the microprocessor (for a quad core Intel™ Nehalem® generation 1 architecture), such as the computational core, the cache memory and I/O bus have been shown on the microprocessor.

## 3. Results and discussions

In case of PMCS, the major challenge is to counter the effects of flow maldistribution induced hot zones, even when thermal load is uniform. Furthermore, flow maldistribution can be minimized and not eradicated as any geometry will possess a finite amount of flow maldistribution. Accordingly, finite temperature non-uniformity is always present on the device irrespective of nature of heat release and this leads to formation of hot zones which are detrimental to device performance. Now, along with the flow maldistribution generated temperature non-uniformity there is an additional challenge that requires addressing. Heat generated by such micro devices is never uniform (due to architecture of the involved circuitry and its preferential activity as required by the system hardware or software) and the thermal load on the PMCS is itself non-uniform, which further leads to worsened cooling. Fig. 2 illustrates the performance of a PMCS for three basic flow distribution configurations (U, I and Z) when a single heater is active, i.e., active heater emits 10W and rest all 2.5 W. The power data has been extracted from a real time i7–4770 3.40GHz quad core processor working at a net load of 70 % its rated power but for a preferentially active processor core out of the four present. The arrangement of the nine heaters also been shown in the figure in accordance to configuration in order to apply the mimicked non uniform heat load. The figures have been plotted for same temperature ranges in order to give a clear picture of comparison of performance of three configurations when non-uniform heat load is present. The figure illustrates the hotspot core temperature (HST), mean temperature of the domain for each case and standard deviation of temperature within domain (representing the uniformity of cooling which is an important factor for overall thermal safety of the device).

As observed from Fig. 2(U), which represents the performance of U configuration, in general, the hot zone temperatures are higher than that of the uniform heat load condition and the temperature non-uniformity in the device is large. From the figure it can be inferred that if any one of the heaters (provided in inset) among (3,1), (3,2), and (3,3) is active, the temperature of the hotspot shoots up and is ~ 20 °C higher than the hotspot temperature in case of uniform heat load. In the event a physical core is situated at such a location, cooling can be a massive challenge during peak performance. The temperatures have been observed to overshoot 70 °C in several occasions and this can lead to device failure if the processing load increases. It can be concluded that for the considered flow configuration (i.e. U), if the heat generation by the device is more at any of the 3 discussed locations, there is high probability of thermal failure of the device. However, the high standard deviation of temperature distribution in the system is also a quantitative indicator of the changed morphology of the hot spot compared to uniform case. In case of U configuration, if any of the regions except (3, 1), (3, 2), and (3, 3) are analogous to an active core and generates higher heat flux than the rest, it can be considered nearly similar to the uniform flux case and it can be said that the device is thermally as safe as uniform case, with some additional tolerance.

Fig. 2(I) represents the performance of I configuration. From the figure it can be inferred that the maximum temperatures for non-uniform thermal load case are always appreciably higher compared to uniform case. The hot spot temperature shoots up to ~ 17 $^{o}$C higher than maximum temperature in case of uniform load. However, the performance of I configuration is not as bad as U configuration since in I the flow maldistribution is less compared to U (Maganti et al. [17]). Hence, the effect of flow maldistribution on temperature maldistribution is already less in case of I compared to U. Maximum hot spot temperature is obtained if the location of active heater is at any of the regions, (1,1); (1,2); (1,3); (3,1); (3,2) and (3,3) and there are significant probabilities of thermal failure of the device. Fig. 2 (Z) represents the thermal performance of Z configuration for non-uniform heat load. From the figure it can be observed that the maximum temperature rises by ~ 14 $^{o}$C in case of non-uniform compared to uniform case. However, it can be observed from the figure that the magnitude of maximum temperature shows weak dependence on the location of the active heater. Since Z has even lower flow maldistribution when compared with I, hence the effect of the same on thermal maldistribution is further reduced. Accordingly, cooling systems of any configuration (U, I and Z), designed employing assumptions with shortcomings (i.e. thermal load to PMCS is uniform) would in really be ineffective and the device would fail, even possibly beyond repair or recovery. It is deemed essential that the probabilistic location(s) of active heater(s) be determined beforehand for a device in order to implement the best strategic cooling system.

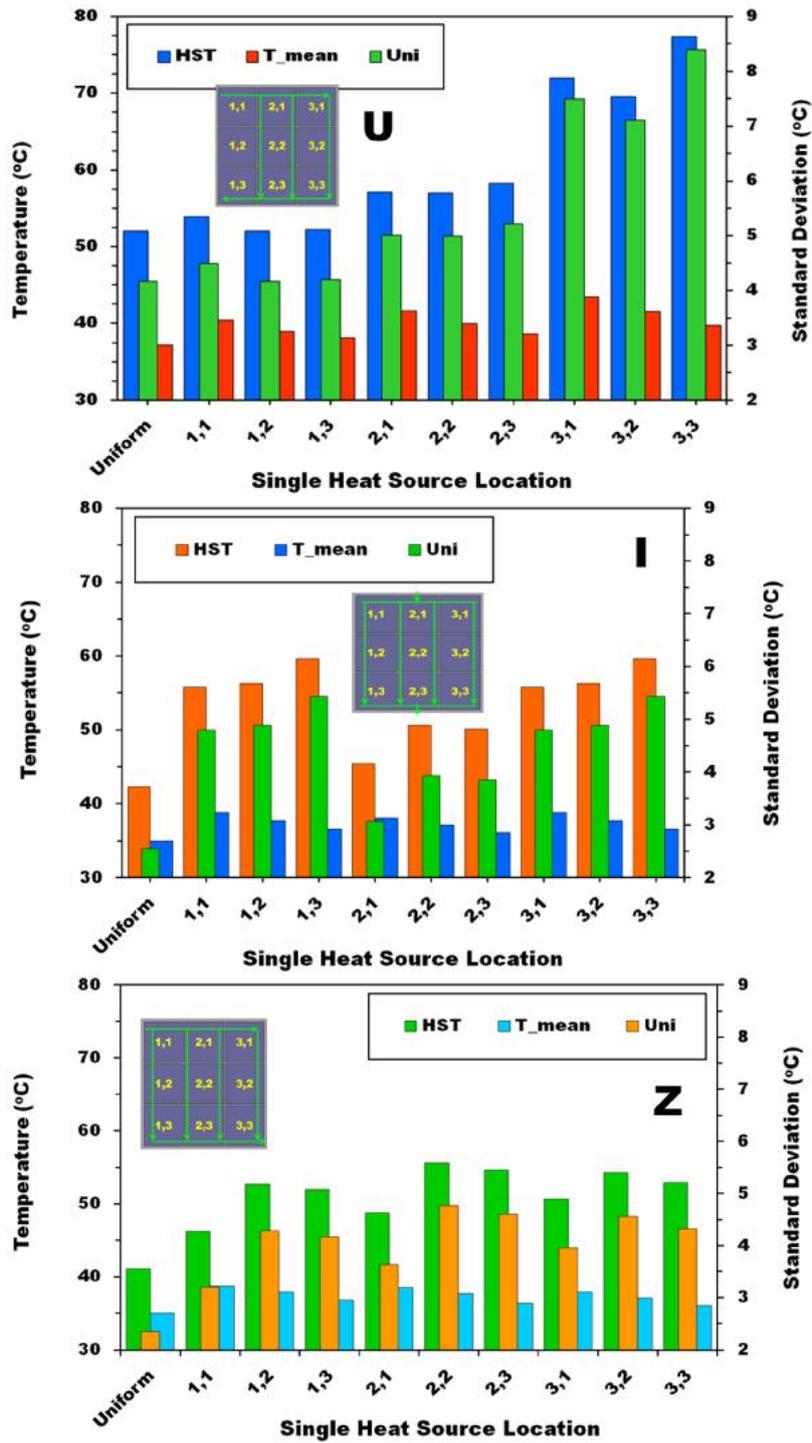

**FIGURE 2:** Effects of non-uniform thermal load on the performance of PMCS (active heater at 10W and rest at 2.5 W, as mapped from a real time microprocessor).

In addition to increase in hot spot core temperature, the shape and spread of the same as well as the location changes for different active heater positions. Such observations have been qualitatively shown for U, I and Z in Figs. 3, 4 and 5 respectively, which illustrate the thermal contours for uniform and non-uniform thermal load cases. Figures (a) represent uniform heat flux and (b), (c), (d), (e), (f), (g), (h), (i) and (j) represent active heaters in positions (1,1), (1,2), (1,3), (2,1), (2,2), (2,3), (3,1), (3,2) and (3,3) respectively. From the figure it can be observed that the location, shape and size of hotspots are no longer same as uniform thermal load case. In case of uniform thermal load, the sole criterion determining the location of the hotspot is the nature of flow maldistribution within the device. For a given configuration, the location of hot spot and shape remains same in case of uniform thermal load if the thermal load remains same. Whereas in case of non-uniform thermal load, it is the combined effect of both flow maldistribution and non-uniform thermal load which decides the characteristics of the resultant hot spot. From Fig. 3 (which represents U configuration contours) it can be inferred that when non-uniform heat load is applied, there is not much change in location of hotspot compared with uniform case in most of the cases studied. Since U configuration has high maldistribution compared to I and Z, the location of the hot spot is largely influenced by flow maldistribution and weakly by the non-uniformity of heat load. Accordingly, for most of the cases, location of hot spot is similar to uniform thermal load, however, the spread and core temperatures are often augmented. It is of interest to observe that in few cases, more than one hot spot erupt and this in fact establishes the proposal that uniform heat load assumptions can lead to drastic miscalculation of cooling system requirements.

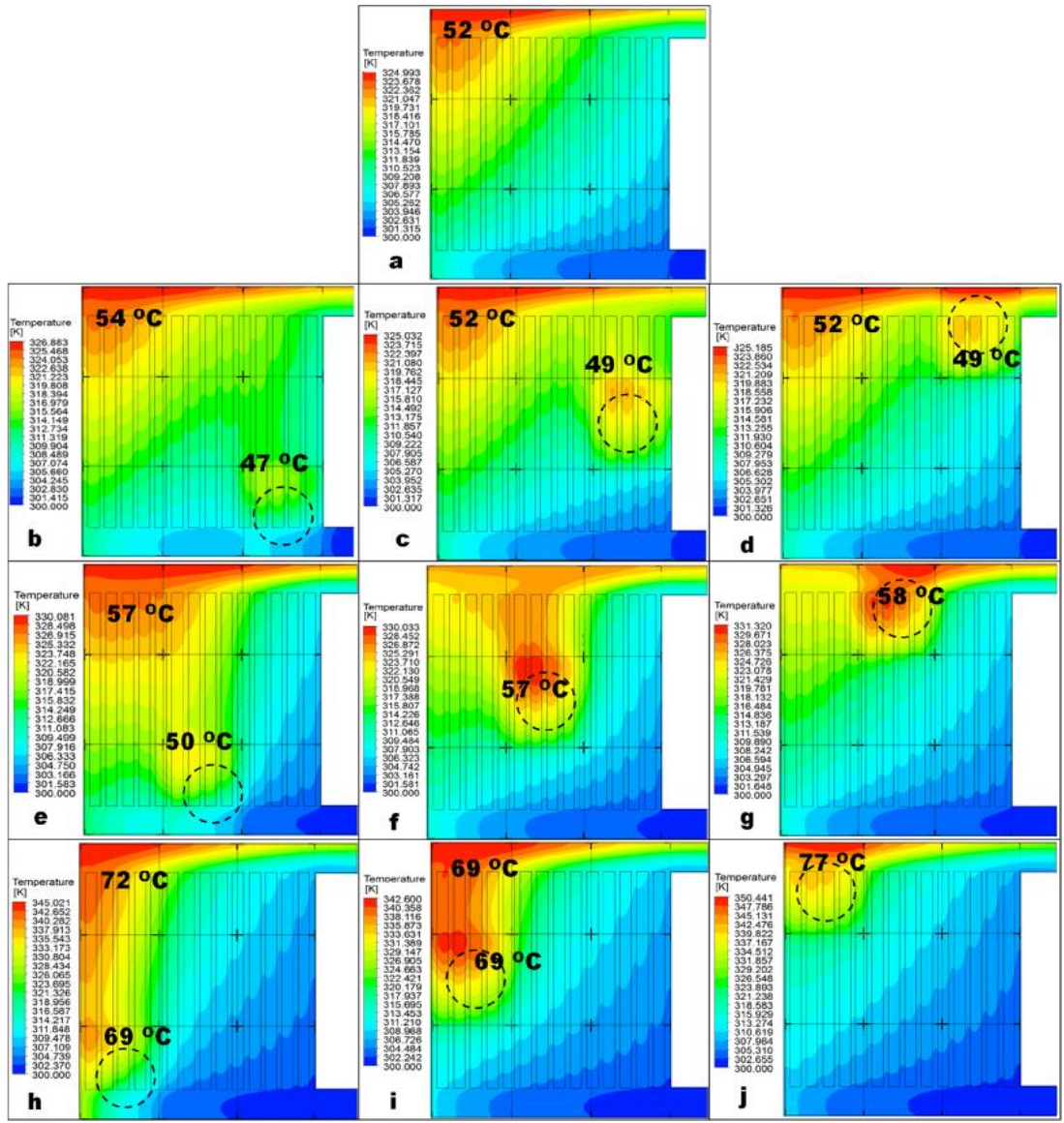

**FIGURE 3:** Qualitative representation of temperature patterns and hot spot locations for both uniform and non-uniform thermal load (single heater) in U configuration. The dotted circle represents the peak load heater position.

Whereas in case of I and Z configurations, the location of hot spot no longer follows uniform thermal load case and it is due to relatively low flow maldistribution compared with U. From Figs 3 and 4 it can be observed that depending on active heater location, the location of hot spot changes. It can also be concluded that the maximum temperature within device strongly depends on active heater location even when configuration of PMCS remains same. The spread of the hot spot is much pronounced in several cases of non-uniform heat load. It is also to be noted that in some cases, the core temperature of hot spot for U configuration may be higher but the size of hotspot is confined to smaller areas compared to I and Z and it is due to the flow patterns in respective configurations. However, as the spot temperature in U are generally higher, the smaller spot size necessarily does not imply favourable conditions.

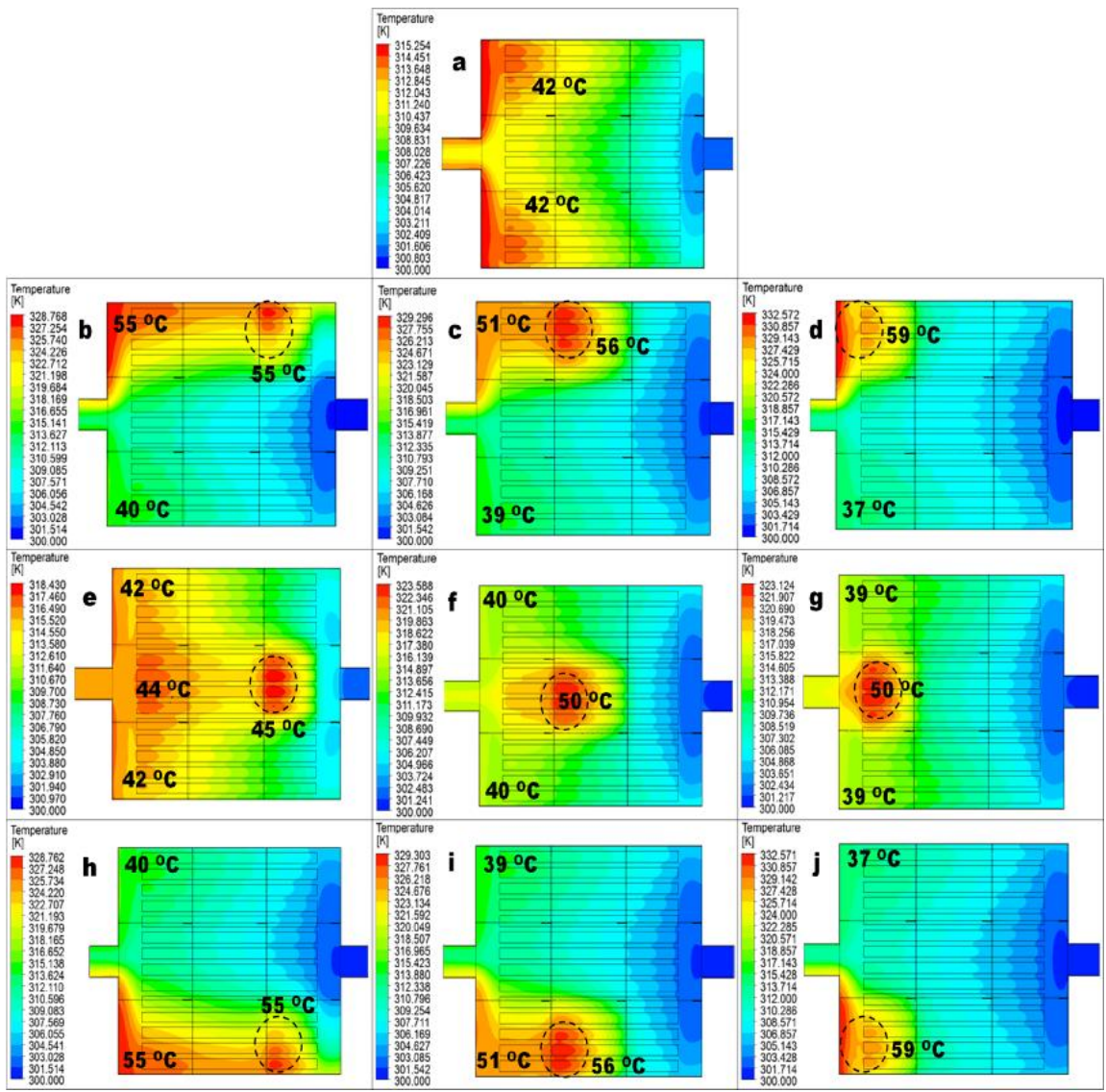

**FIGURE 4:** Qualitative representation of temperature patterns and hot spot locations for both uniform and non-uniform thermal load (single heater) in I configuration. The dotted circle represents the peak load heater position.

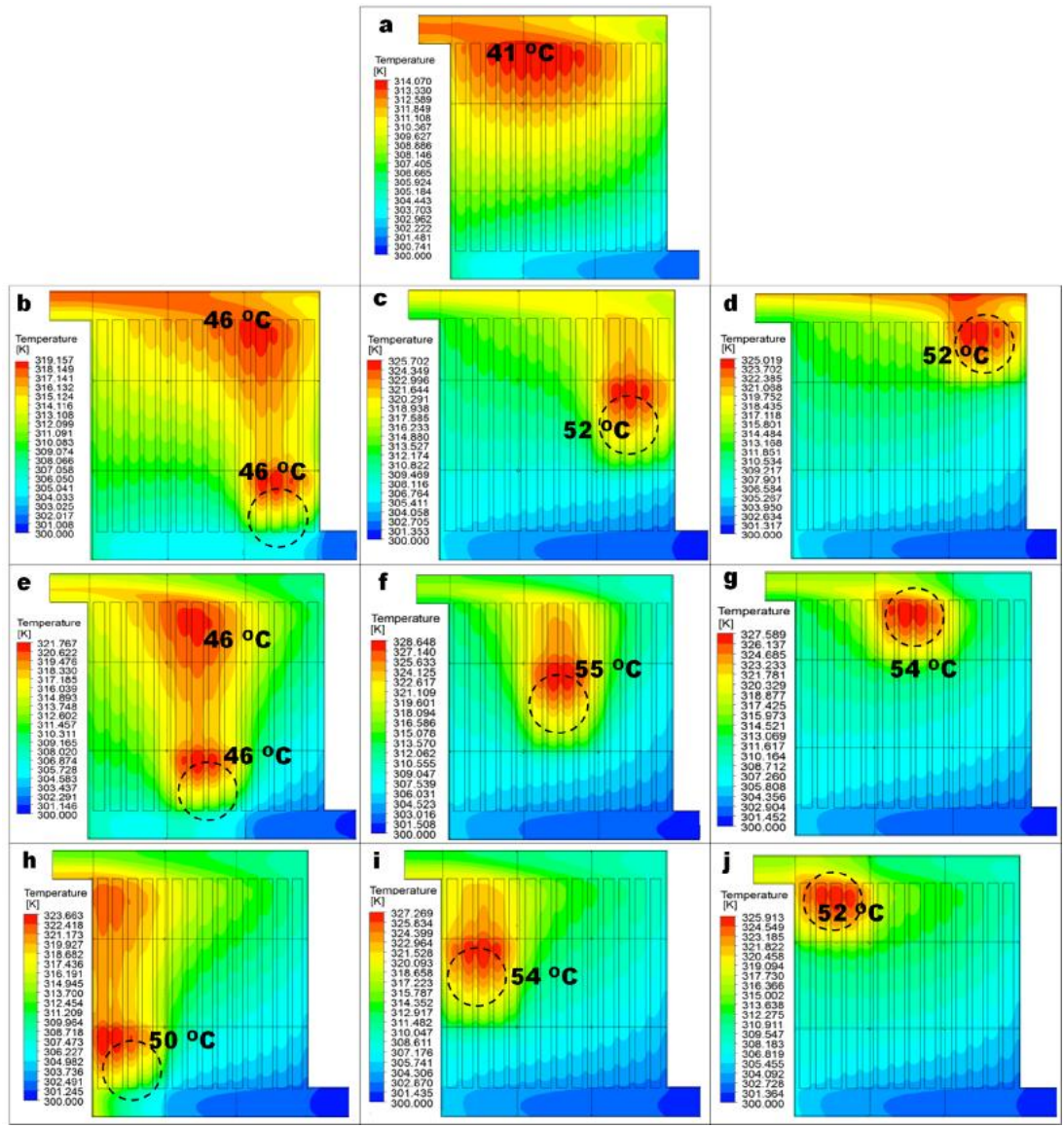

**FIGURE 5:** Qualitative representation of temperature patterns and hot spot locations for both uniform and non-uniform thermal load (single heater) in Z configuration. The dotted circle represents the peak load heater position.

As and when such real systems are involved, higher cooling is often desirable. Employing suitable, thermally high efficiency fluids can address such problems. Employing nanofluids in such high performance cooling devices will not only serve the purpose of more cooling but also improve the uniformity of cooling. Fig. 4 illustrates comparison of thermal performance of water and alumina-water (5 vol. %) nanofluids as working fluids in PMCS where hot spot temperature (i.e. HST) in case of both water and nanofluid and uniformity of cooling (i.e. standard deviation of temperature within domain) are shown. From Fig. 6(a) it can be inferred that the core temperature of hot spot for uniform, (1, 1), (1, 2), (1, 3), (2, 1), (2, 2) and (2, 3) are more or less same and magnitude of maximum temperature is predominantly less compared with (3, 1), (3, 2) and (3, 3) as active heater locations. In case of uniform, (1, 1), (1, 2), (1, 3), (2, 1), (2, 2) and (2, 3), difference between core temperatures of water and nanofluids is same. However, the difference is enhanced in case of (3, 1), (3, 2) and (3, 3). Since temperature is high in these cases, slip mechanisms like Brownian and thermophoresis will be more active, such that more random motion of particles within flow domain leads to augmented heat transfer, resulting in higher cooling. In similitude to reports by present authors (Maganti et al. [10] and Maganti et al. [17]), the smart nature of nanofluids has been established again i.e., at high temperatures thermal performance of nanofluids is higher. The same is seen from Fig. 6 (U) (a) where percentage decrease in maximum temperature when nanofluid is used as working fluid has been shown. It can be observed that the decrease in core temperature of hot spot is more in case of (3, 1), (3, 2) and (3, 3) active heater locations. It can also be inferred that uniformity of cooling has improved along with reduced hot spot core temperature which again proves the '*smart*' fluid effect. However, such pronounced smart effects have not been observed in case of I and Z. As discussed, the flow distribution is uniform in I and Z compared to U and it is due to the arrangement of channels with respect to inlet and outlet manifolds. Since the Reynolds number is high (Re=300), due to high flow inertia, resistance to random motion of particles is high (Maganti et al. [17]). Accordingly, high temperatures are required to active the phenomena like appreciable Brownian diffusion and thermophoresis to overcome the inertial resistance to behave as smart fluid. From the figure it can be observed that the maximum temperature in case of U configuration shoots up to 80 $^0$C, whereas in case of I and Z it is ~ 50 $^0$C. Hence the smart effect can be seen in case of U configuration when (3, 1), (3, 2) and (3, 3) are active heaters.

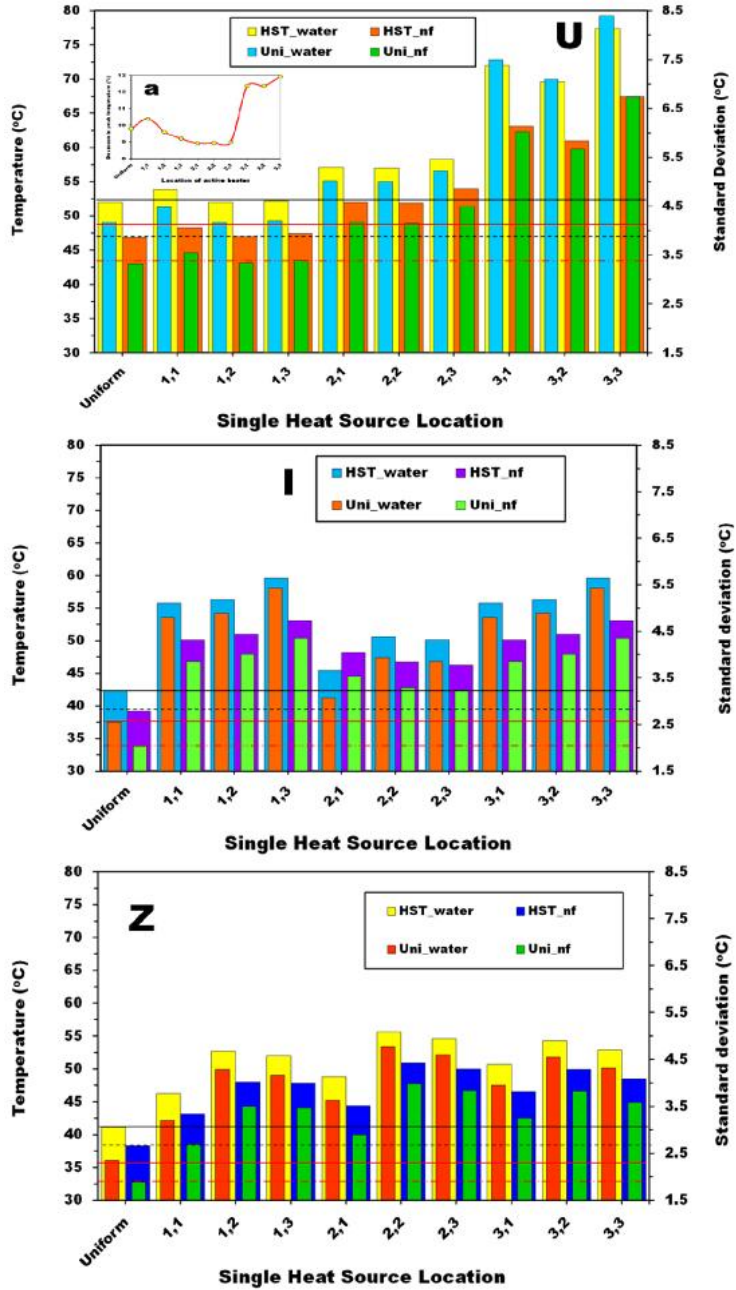

**FIGURE 6:** Comparison of thermal performance of water and alumina-water (5 vol. %) nanofluids as heat transfer fluids in parallel microchannel cooling systems under non-uniform thermal load.

U configuration exhibits highest flow maldistribution, followed by I and least for Z. In case of U, the arrangement of channels with respect to inlet and outlet manifold are such that former channels are flooded with fluid whereas later channels are starved. Due to extremely non uniform distribution of fluid, U is often not recommended for cooling electronic devices due to generation of high hot spot core temperatures even for uniform heat loads. Since I and Z are having better flow distribution compared to U, those are highly recommended configurations for cooing microelectronic devices if thermal load is uniform. However, in realistic cases (thermal load to PMCS is inherently non-uniform), the common intuition that U is to be avoided for PMCS, is invalidated. According to the flow patterns of U, since former channels get more mass flux of coolant, it is to be preferred as the best configuration for cooling electronic devices if the active region coincides with the former channels. The suitable and best configuration(s) depending on active heater location have been tabulated in Figure. 7 and this selection have been performed based on qualitative analysis of the contours and quantitatively based on a proposed *Figure of Merit* (Maganti et al. [17]) (which quantifies the thermal performance and uniformity of cooling of a given configuration for a set of specified working conditions). From the figure it can be inferred that the most uniform flow distribution case (Z configuration) is essentially a poor performance system if active heat source locations are present at any of the places among (2, 1), (2, 2) or (2, 3). On the contrary, the most maldistributed case (U configuration) is the most desirable if active heat source is present in any of locations among (1, 2) and (1, 3). From fig. 7 it can be concluded that flow maldistribution in reality is not always a bane, but can help to tackle cooling challenges if thermal loads are non-uniform. However, thermal load morphology and distribution should be estimated in order to propose suitable configuration for better cooling of microelectronic devices. Hence, the present study conclusively shows that maldistribution can be engineered to cool hot spots efficiently. Furthermore, a U type system can be accordingly modified to cater to all cooling challenges as from point of view of fluid infusion and collection; the U system stands most effective due to its inlet and outlet manifolds directed on the same side of the device.

| Active Hotspot Location | U | I | Z |
|---|---|---|---|
| (center-right middle row, all cells) | ✗ 2.08 | ✓ 2.76 | ✓ 2.91 |
| top-right | ✗ 2.00 | ✗ 1.93 | ✓ 2.40 |
| top-center | ✓ 2.08 | ✗ 1.92 | ✓ 2.05 |
| top-left | ✓ 2.07 | ✗ 1.82 | ✓ 2.08 |
| middle-right | ✗ 1.89 | ✓ 2.25 | ✗ 2.18 |
| middle-center | ✗ 1.90 | ✓ 2.14 | ✗ 1.94 |
| middle-left | ✗ 1.86 | ✓ 2.16 | ✗ 1.97 |
| bottom-right | ✗ 1.60 | ✗ 1.93 | ✓ 2.13 |
| bottom-center | ✗ 1.63 | ✓ 1.92 | ✓ 1.98 |
| bottom-left | ✗ 1.53 | ✗ 1.82 | ✓ 2.04 |

**FIGURE 7:** Proposed flow configuration(s) for thermal safety of device based on location of active heat source (single active heater). The values represent the corresponding *Figure of Merit*.

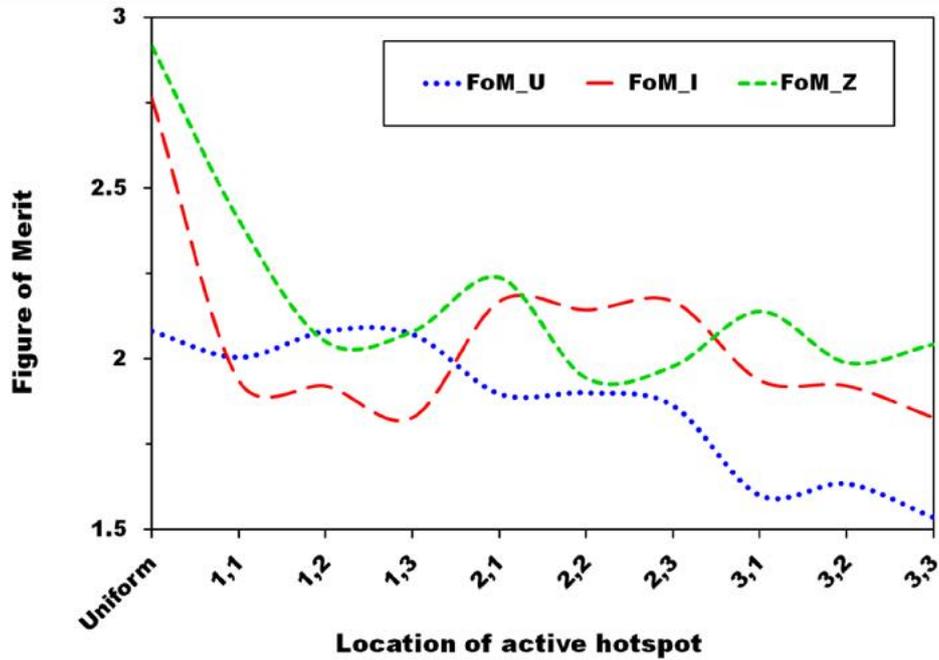

**FIGURE 8:** Figure of Merit of each configuration for known working parameters.

A single hot spot is also an idealized condition for a real microprocessor where the physical cores are often active in groups. Having shown the performance of PMCS when single heater is active, Fig. 9 illustrates thermal performance of U, I and Z configurations when two hot spots are active (i.e. active hot spots at 10W and rest all at 2.5 W). The locations of heaters and thermal load have been applied by mimicking real time data of an Intel® Core™ i7–4770 3.40 GHz processor. According to the architecture of the said processor (as shown in the Fig. 1), the location of heaters considered closely overlaps with the physical cores. The figure has been plotted for hot spot core temperature (HST) and standard deviation of temperature (represented by '*Uni*') within domain. From Fig. 9 (U) it can be inferred that U configuration PMCS is recommendable for such architecture and the performance of such cooling system will not change much by changing location of active heaters. Since the heater locations are coincident with the former channels, thermal performance of U configuration can be said to be satisfactory for given non uniform thermal load. However, if further cooling is desired, employing nanofluids instead of simple fluids seem to solve the problem to a large extent. Fig. 9 (I) has been shown for I configuration and it can be inferred that thermal performance of I and U configurations are more or less same and both of them show reasonably satisfactory performance. Fig. 9 (Z) illustrates that thermal performance of Z configuration is remarkably high compared with U and I configurations. It is due to the fact that Z configuration having relatively better flow distribution so that effect of same on cooling performance is more which leads to more uniform cooling even when non uniform thermal load applied.

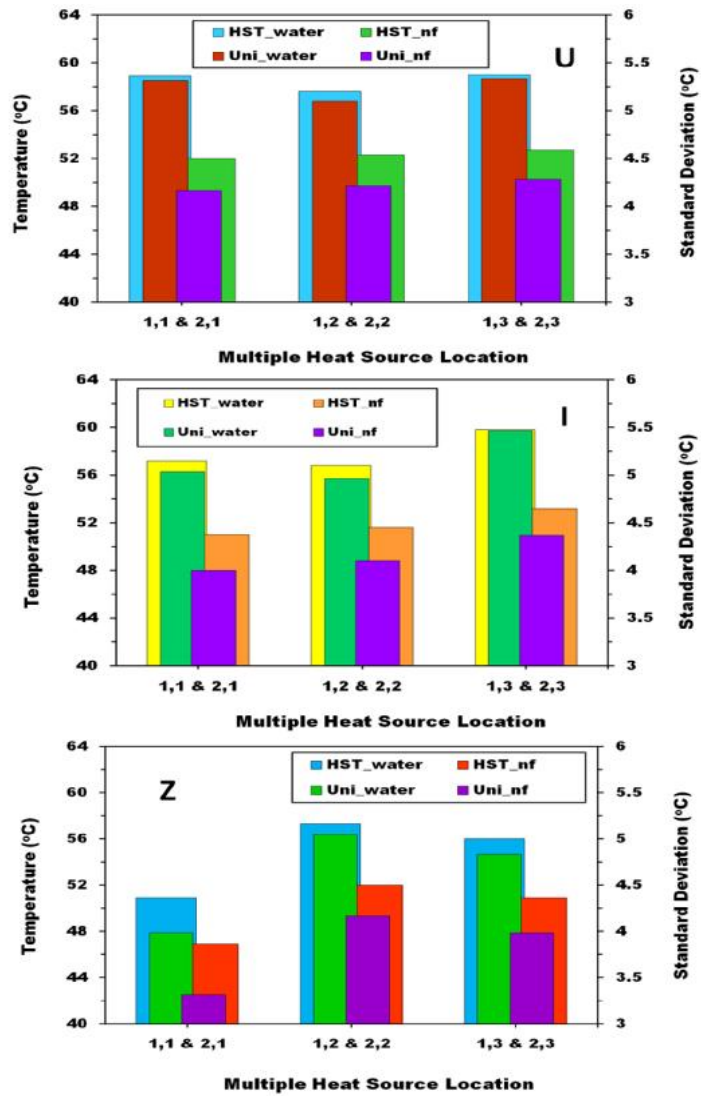

**FIGURE 9:** Effect of non-uniform thermal load on thermal performance of water and alumina-water based PMCS when two heaters (simulating a real processor physical core) are active.

Even though U and I systems are providing moderately satisfactory cooling when two heaters are active, Fig. 10 illustrates proposed configurations based on superior thermal performance for a given pair of active heaters. From the figure it can be inferred that if location of an active heater pair are at (1, 1) & (2, 1), Z is the most suitable between I and Z configurations. If locations are at (1, 2) & (2, 2), all three configurations can be for thermal management. Whereas, again Z is best suited if location of active heaters are at (1, 3) & (2, 3). Having said that, for a given processor architecture, i.e. Intel® Core™ i7–4770 3.40 GHz, any of the three basic configurations (U, I and Z) can be used for moderately agreeable thermal performance if the location of hot spot can be estimated *a priori*. With information about the location of an active heat source, an optimal configuration for mitigating the resulting hot spot can be obtained from the present article. It is also noteworthy that the usage of nanofluids also results in reduction in the hot spot core area simultaneously with enhanced cooling (Fig. 11 (c)) is a result of employing the DPM formulation. The thermo-fluidic slip mechanisms ensure that more heat is extracted from a hot spot due to its higher temperature. The effective property formulations are unable to capture such phenomena due to its inability to behave in accordance to local thermal conditions. Hence, nanofluids not only reduce the hot spot temperatures, but are also '*smart*' enough to shrink the hot spot size thereby ensuring higher safety.

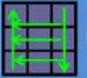

**FIGURE 10:** Proposed cooling configurations depending on location of active heat sources (two active heaters) mimicking the physical core of an Intel® Core™ i7–4770 3.40 GHz processor.

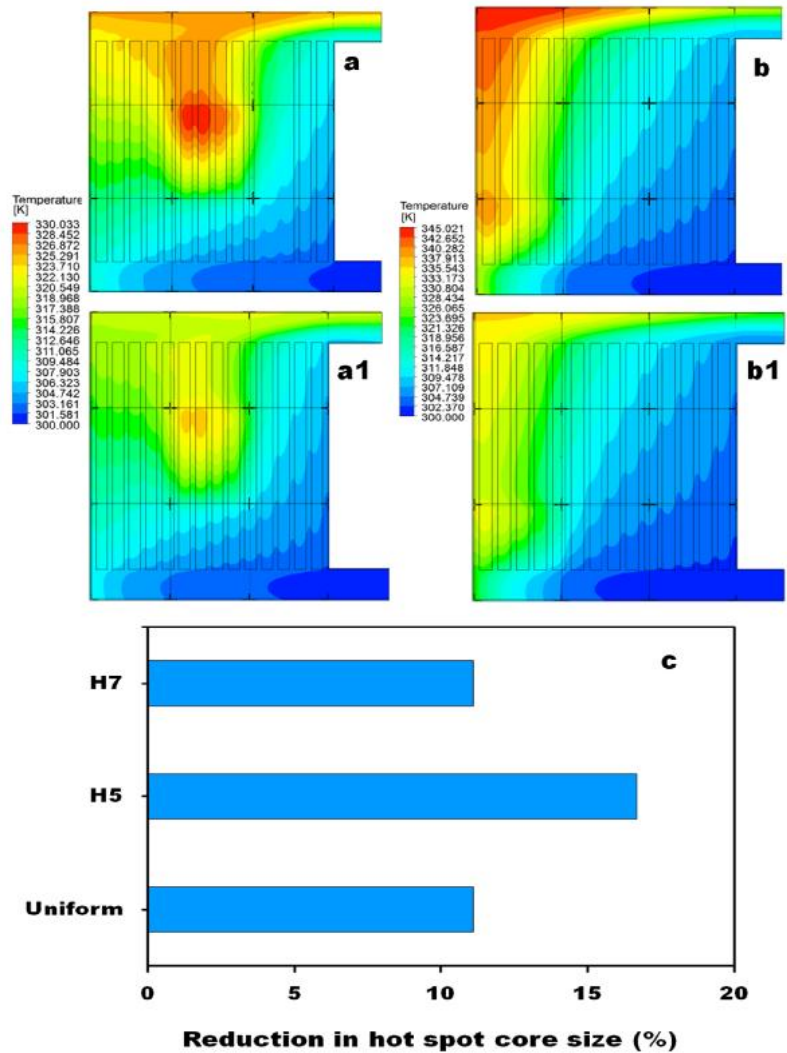

**FIGURE 11:** Thermal contours of U configuration using water and nanofluids as working fluids, (a) and (a1) when H5 heater is active (b) and (b1) when H7 heater is active. The reduction in temperature as well as hot spot size is markedly visible. (c) Percentage reduction in hot spot core size using nanofluids with respect to water as working fluids.

## 4. CONCLUSIONS

The additional effects of non-uniform heat release from an Intel i7 quad core microprocessor on the thermal maldistribution in a PMCS in addition to flow maldistribution induced thermal maldistribution have been reported. The present study focuses on understanding the implications of such thermal maldistribution and cooling strategies using such a PMCS have been discussed. Based on the architecture of an Intel® Core™ i7–4770 3.40 GHz microprocessor (Nelahem architecture), a 3 x 3 heater array has been implemented to induce non-uniform heat generation. It has been observed that in case of U, I and Z configurations, the hotspot core temperature can go up as high as ~ 20 $^{o}$C, 17 $^{o}$C and 15 $^{o}$C respectively more than hotspot core temperatures in case of uniform heat load assumption when single heat source is active. Furthermore, the hotspot locations as well as size of the spot changes considerably and device safety is compromised in several cases. It has also been found that in case of real i7 processors, when a core is active, in reality 2 heaters need to be active for mimicking the effects. It has been found that water is not an efficient coolant for uniform cooling in several instances. Accordingly, it has been revealed that nanofluids are very efficient to bring about uniform and as well as increased cooling in such PMCS. Slip mechanisms like Brownian and thermophoresis are found to be reason behind more uniform cooling at high temperatures, leading to shrinking of hot spot size and ensuring device safety. Suitable configurations have been proposed for mitigating non uniform thermal load induced hotspots for a known thermal load distribution. The present work is important to understand the thermal effects within a real time microprocessor and provides a foundation for criteria of selection of the suitable PMCS for optimal performance and highest safety of device.

## Acknowledgements

LSM would like to thank the Ministry of Human Resource Development (Govt. of India) for the doctoral scholarship and IIT Ropar for facilitating partial support to the research work.